\documentclass[a4paper, 10pt, conference]{ieeeconf}      
\IEEEoverridecommandlockouts                              
\usepackage{cite}
\usepackage{amsmath,amssymb,amsfonts,bm}
\usepackage{graphicx}
\usepackage{textcomp}
\usepackage{tikz}
\usepackage{subfig}
\usepackage{acronym}
\usepackage{multicol}
\usepackage{upgreek}
\usepackage{tcolorbox}
\usepackage{stfloats}
\usepackage{upgreek}
\usepackage[percent]{overpic}
\usepackage{balance}

\tikzstyle{fontbf} = [font=\bf]
\usetikzlibrary{arrows, calc,
decorations.pathmorphing,decorations.pathreplacing, decorations.markings,backgrounds, positioning,
fit, shapes.geometric, patterns}

\tikzstyle{rect} = [rectangle, minimum width=3cm, minimum height=1cm, text centered, draw=black]

\newcommand{\n}{\noindent}

\renewcommand{\vec}[1]{{\bm{#1}}}
\newcommand{\mat}[1]{{\bm{#1}}}

\newcommand{\blc}[1]{{\tilde{#1}}}
\newcommand{\rem}[1]{{\bar{#1}}}

\newcommand{\cvec}[1]{{\bm{\mathsf{#1}}}}
\newcommand{\cmat}[1]{{\bm{\mathsf{#1}}}}

\newcommand{\appref}[1]{Appendix}

\newcommand{\figref}[1]{Fig.~\ref{#1}}

\newcommand{\secref}[1]{Section~\ref{#1}}

\acrodef{ASM}{Active Set Methods}
\acrodef{DARE}{Discrete-time Algebraic Riccati Equation}
\acrodef{GP}{Gradient Projection}
\acrodef{CG}{Conjugate Gradient}
\acrodef{FGM}{Fast Gradient Methods}
\acrodef{FLOPs}{Floating Point Operations}
\acrodef{IPM}{Interior Point Methods}
\acrodef{KKT}{Karush--Kuhn--Tucker}
\acrodef{LQR}{Linear Quadratic Regulator}
\acrodef{LTV}{Linear Time-Varying}
\acrodef{MB}{Move Blocking}
\acrodef{MHE}{Moving Horizon Estimation}
\acrodef{MINRES}{MINimum RESidual}
\acrodef{MPC}{Model Predictive Control}
\acrodef{MSS}{Multi-Stage Separation}
\acrodef{NLP}{Nonlinear Programming}
\acrodef{NPP}{Newton Projection with Proportioning}
\acrodef{OCP}{Optimal Control Problem}
\acrodef{PLS}{Projected Line Search}
\acrodef{PD}{Positive Definite}
\acrodef{PSD}{Positive Semi-definite}
\acrodef{QP}{Quadratic Programming}
\acrodef{RAM}{Random Access Memory}
\acrodef{RHC}{Receding Horizon Control}
\acrodef{NLP}{NonLinear Programming}
\acrodef{SQP}{Sequential Quadratic Programming}
\acrodef{NNZs}{Number of non-zero elements}

\title{\LARGE \bf
On the Quadratic Programming Solution for\\Model Predictive Control with Move Blocking
}

\author{Pavel~Otta$^{1}$, Ond\v{r}ej~\v{S}antin$^{2}$, and Vladim\'{i}r Havlena$^{1}$
\thanks{The first author was supported by SGS (the Student Grant Fund) of the Czech Technical University.}
\thanks{$^{1}$Czech Technical University in Prague, Technick\'{a} 2, CZ-16627 Prague 6, Czech Republic
        {\tt\small {ottapav1, havlena}@fel.cvut.cz}}%
\thanks{$^{2}$Garrett Motion Czech Republic s.r.o., Tu\v{r}anka 1387/100, CZ-62700 Brno - Slatina, Czech Republic
        {\tt\small ondrej.santin@garrettmotion.com}}%
}

\begin{document}

\maketitle

\begin{abstract}
Model Predictive Control (MPC) is a popular optimization-based control technique. MPC is usually formulated as sparse or dense Quadratic Programming (QP). This paper reviews two well-known methods, namely, state condensing and move blocking, and brings them together. Their combination results in generalized QP that serves arbitrarily sparse (or dense) QP for MPC with \emph{move blocking}. The proposed QP can be solved by a specialized solver capable of exploiting a sparsity structure of the problem. Numerical examples give inside in computational and memory requirements.
\end{abstract}


\section{Introduction}
\label{sec:introduction}
\ac{MPC} is a popular multivariable control technique that systematically incorporates physical constraints (e.g., potential or flow limits) by design. It is an optimization-based method, i.e., in every sampling period, a finite optimal control problem needs to be solved. For linear dynamics and quadratic costs, the problem to be solved is structured \ac{QP}. For nonlinear dynamics and nonlinear costs, the problem to be solved is \ac{NLP}. \ac{NLP} can be solved by \ac{SQP}, which requires a sequence of structured \ac{QP} to be solved. Therefore, the results presented in this paper can address nonlinear MPC as well.

The two most common \ac{QP} formulations in which the \ac{MPC} problem can be written are the \emph{dense} and the \emph{sparse} formulation \cite{Jerez2011}. In the sparse \ac{QP} formulation, the minimization variables are inputs and states over the prediction horizon, and they are interconnected by the equality constraint representing system dynamics explicitly. Consequently, the \ac{QP} problem is large in a number of variables with a specific sparsity pattern. On the other hand, in the dense \ac{QP} formulation, the minimization variables are inputs over the prediction horizon only, the states are eliminated out, and the interconnection is held implicitly. Consequently, the \ac{QP} problem is smaller, but with no sparsity pattern.

A comparison between sparse and dense \ac{QP} formulation in the context of walking motion generation is presented in \cite{Dimitrov2011}. This application benefits from the use of sparse \ac{QP} formulation as parameters change in their model; only a negligible additional computational effort is required. In some other applications (e.g., \cite{Richter2010}), it might be beneficial to transform sparse QP to the dense one by a so-called \emph{condensing} procedure. When a model is fixed, the condensing can be performed once offline, which leads to a significant computational saving. Note that condensing can be done with quadratic complexity in horizon length, as proposed in \cite{Frison2016}.

It should be noted that dense \ac{QP} can be solved by a generic-purpose solver, which made the dense formulation more popular in the past. Nowadays, several structure-exploiting methods tailored for sparse \ac{QP} arising in \ac{MPC} exist (e.g. \cite{Wang2008,Frasch2014,NPPsparse,Ullmann2011}). As the problem formulation proposed in this work is derived from the sparse \ac{QP}, these algorithms can be applied with no additional modification required.

When treating dense formulation, the computation complexity can be decreased by the \emph{move-blocking} technique. The idea of this technique is to fix consequent inputs at the same value. Therefore the number of degrees of freedom of the optimization problem decreases. Another strategy to reduce the number of degrees of freedom is to utilize \ac{LQR} \cite{Cagienard2007}. A move-blocking strategy can optimize control performance, robustness, or feasibility when hard state constraints are considered \cite{Shekhar2012,Shekhar2015,Gondhalekar2010}. The idea of using move blocking regardless of the sampling period was proposed in \cite{Faroni2017}.

State condensing has been proposed in \cite{Axehill2015}. It enables sparsity level control of the \ac{QP} formulation. The level of sparsity can be controlled smoothly in between the sparse (non-condensed) to dense (fully-condensed) QP formulation. This method was used for sped-up dual Newton step algorithm regarding nonlinear \ac{MPC} in \cite{Kouzoupis2015} or combined with the partial sensitivity update in \cite{Chen2018} recently.

To the best author's knowledge, a generalization of the state condensing for \ac{MPC} problem with move blocking has not been reported in the literature. From the other way around, move blocking has not been adapted for the sparse \ac{QP} formulation yet. This paper attempts to address this issue.





The rest of the paper is organized as follows. In \secref{sec:Model_Predictive_Control}, the \ac{MPC} problem is introduced. \secref{sec:Problem_Statement} gives a basic \ac{QP} problem formulation of the \ac{MPC}. In \secref{sec:Move-Blocking} and \secref{sec:Partial-Condensing}, move-blocking and partial-condensing procedures are described, respectively. \secref{sec:Customization} presents a QP transformation combining state condensing with move blocking. \secref{sec:Numerical_Study} gives an insight in computational and memory requirements of the proposed method based on simulations.

\section{Model Predictive Control}
\label{sec:Model_Predictive_Control}
We are concerned in a discrete-time \ac{LTV} systems in the form
\begin{equation*}\label{eq:NLDTsystem}
\vec{x}(t+1) = \mat{A}(t) \vec{x}(t) + \mat{B}(t)\vec{u}(t) + \vec{w}(t),\quad t\geq 0,
\end{equation*}
where $\mat{A}(t)\in\mathbb{R}^{n_\vec{x}\times n_\vec{x}}$, $\mat{B}(t)\in\mathbb{R}^{n_\vec{x}\times n_\vec{u}}$, and $\vec{w}(t)\in\mathbb{R}^{n_\vec{x}}$ at every time $t$ are known. $\mat{A}(t),\mat{B}(t),\vec{w}(t)$ are deterministic, possibly arising from the linearization of nonlinear system. Input and state dimensions are denoted by $n_\vec{u}$ and $n_\vec{x}$, respectively.


Then the problem of the regulator with input box constraints can be treated as the following \ac{LTV} \ac{MPC}

\n\resizebox{0.95\linewidth}{!}{
\begin{minipage}{\linewidth}
\begin{subequations}\label{eq:MPC_problem}
\begin{eqnarray}\label{eq:MPC_problem_cost}
& \underset{\substack{\vec{u}_{0},\ldots,\vec{u}_{N-1}\\ \vec{x}_{1},\ldots,\vec{x}_{N}}} {\min}&
\dfrac{1}{2} \displaystyle{\sum_{k=0}^{N-1}
\begin{bmatrix}
\vec{u}_{k} \\
\vec{x}_{k}
\end{bmatrix}^T
\begin{bmatrix}
\mat{R}_{k} & \mat{S}^T_{k} \\
\mat{S}_{k} & \mat{Q}_{k}
\end{bmatrix}
\begin{bmatrix}
\vec{u}_{k} \\
\vec{x}_{k}
\end{bmatrix}}
+
\begin{bmatrix}
\vec{u}_{k} \\
\vec{x}_{k}
\end{bmatrix}^T
\begin{bmatrix}
\mat{r}_{k} \\
\mat{q}_{k} 
\end{bmatrix}
\nonumber
\\
& & \quad + \dfrac{1}{2} \vec{x}^T_{N} \mat{Q}_{N} \vec{x}_{N}\\
&  \textrm{s.t.}  & \vec{x}_{k+1} = \mat{A}_{k} \vec{x}_{k} + \mat{B}_{k} \vec{u}_{k} + \vec{w}_{k}, \label{eq:MPCSystemDynamics} \\
& & \underline{\vec{u}}_k \leq \vec{u}_k \leq \vec{\overline{u}}_k, \; k = 0, \ldots, N-1, \\
& & \vec{x}_0 = \hat{\vec{x}}(t), \\\nonumber
\end{eqnarray}
\end{subequations}
\end{minipage}
}
\n where $\hat{\vec{x}}(t) \in \mathbb{R}^{n_\vec{x}}$ is a current state measurement (estimation) at time $t$, $N \in \mathbb{N}$ is the finite prediction horizon length. For the sake of brevity, subscript $k$ denotes time period from the sampling moment $t$. For example, $\vec{x}_k = \vec{x}(t+k)\in \mathbb{R}^{n_\vec{x}}$ and $\vec{u}_k = \vec{u}(t+k) \in \mathbb{R}^{n_\vec{u}}$ denote the state and input at stage $k$ on prediction horizon, respectively. The quadratic weights are $\mat{R}_k \succ \mat{0}$, $\mat{Q}_k - \mat{S}_k\mat{R}_k^{-1}\mat{S}_k^T \succeq \mat{0}$, and the terminal weight $\mat{Q}_{N} \succeq \mat{0}$. The optimizer of problem \eqref{eq:MPC_problem} is a unique input sequence with associated state trajectory.


As there is new measurement $\hat{\vec{x}}(t)$ and time-varying model $\mat{A}_k,\mat{B}_k,\vec{w}_k, k = 0, \ldots, N-1$ along the prediction horizon is available at each sampling instant, the problem \eqref{eq:MPC_problem} has to be reoptimized. Hence{,} the \emph{receding horizon} concept is established, i.e.{,} the plan of control inputs $\vec{u}_0, \ldots, \vec{u}_{N-1}$ is recomputed at each sampling instant parametrized by measured system state $\hat{\vec{x}}$ and only the first control move $\vec{u}_0$ is actually applied to the system, cf. \cite{Mayne2000}. The need for re-computation requires a fast solver for problem \eqref{eq:MPC_problem}{,} to have a solution ready by the next sampling time.

\section{Sparse QP Formulation}
\label{sec:Problem_Statement}
Problem \eqref{eq:MPC_problem} can be rewritten straightforwardly as a box-constrained \ac{QP} in the following form
\\\n\resizebox{\linewidth}{!}{
\begin{minipage}{\linewidth}
\begin{subequations}\label{eq:QP_problem}
\begin{eqnarray}
 & \underset{\cvec{u},\cvec{x}} {\min}& \dfrac{1}{2}
\begin{bmatrix}
\cvec{u} \\
\cvec{x}
\end{bmatrix}^T
\begin{bmatrix}
\cmat{R} & \cmat{S}^T \\
\cmat{S} & \cmat{Q}
\end{bmatrix}
\begin{bmatrix}
\cvec{u} \\
\cvec{x}
\end{bmatrix}
+
\begin{bmatrix}
\cvec{u} \\
\cvec{x}
\end{bmatrix}^T
\begin{bmatrix}
\cvec{r}(\vec{x}_0) \\
\cvec{q}
\end{bmatrix}\nonumber
\\ & & \quad + c(\vec{x}_0) \label{eq:QP_cost} \\
&  \textrm{s.t.}  & \cmat{A}\cvec{x} = \cmat{B}\cvec{u} + \cvec{w}(\vec{x}_0), \label{eq:QP_equality_constraints} \\
& & \underline{\cmat{u}} \leq \cvec{u} \leq \overline{\cvec{u}}, \label{eq:QP_inequality_constraints}
\end{eqnarray}
\end{subequations}
\end{minipage}
}\\

\n parametrized by $\vec{x}_0\in\mathbb{R}^{n_\vec{x}}$ where $\cvec{u}\in\mathbb{R}^{N\cdot n_\vec{u}}$ denotes box-constrained inputs sequence and $\cvec{x}\in\mathbb{R}^{N\cdot n_\vec{x}}$ denotes states trajectory vectors stacked as
\n\resizebox{\linewidth}{!}{
\begin{minipage}{\linewidth}
\begin{eqnarray*}
\cmat{x} &= 
\begin{bmatrix}
\vec{x}_1^T,\ldots,\vec{x}_{N-1}^T,\vec{x}_{N}^T
\end{bmatrix}^T\!\!,\quad
\cmat{u} = 
\begin{bmatrix}
\vec{u}_0^T,\ldots,\vec{u}_{N-2}^T,\vec{u}_{N-1}^T
\end{bmatrix}^T\!\!
\end{eqnarray*}
\end{minipage}
}

Presented class of problems defined by \eqref{eq:QP_problem} \emph{does impose} box input constraints. It is motivated by the fact that algorithm controls actuators, in common, which operate in a limited range (e.g., valve position) or limited range for rate of change (e.g., valve transition speed). The benefit is that box-constrained \ac{QP} can be solved faster than a generally-constrained one.
The presented approach \emph{does not impose} hard state constraints. It prevents the feasibility issue. However, state limits can be imposed as soft constraints using the penalty method \cite{Kerrigan2000}. 
We believe that the class of problems defined by \eqref{eq:QP_problem} is wide enough to cover the majority of industrial problems. 

The assumption that have been made is
$
\begin{bmatrix}
\cmat{Q} & \cmat{S}^T \\
\cmat{S} & \cmat{R} 
\end{bmatrix}
$ is positive semidefinite. The individual vectors and matrices in \eqref{eq:QP_cost} are composed as follows
\\\n\resizebox{\linewidth}{!}{
\begin{minipage}{\linewidth}
\begin{align*}
\cmat{R}
&=
\begin{bmatrix}
\mat{R}_0 \\
& \mat{R}_1 \\
&& \ddots                     \\
&&& \mat{R}_{N-1}           \\
\end{bmatrix}\!\!,~
\cmat{r}(\mat{x}_0) = 
\begin{bmatrix}
\mat{S}_0\vec{x}_0 + \vec{r}_0 \\
\vdots \\
\vec{r}_{N-2} \\
\vec{r}_{N-1}
\end{bmatrix}\!\!,~
\cmat{q} = 
\begin{bmatrix}
\vec{q}_1 \\
\vdots \\
\vec{q}_{N-1} \\
\vec{q}_{N}
\end{bmatrix}\!\!,
\\
\cmat{S} &=
\begin{bmatrix}
\mat{0} & \mat{S}_1         \\
&& \ddots        \\
&&& \mat{S}_{N-1}   \\
&&& \mat{0}           \\
\end{bmatrix}\!\!,\quad
\cmat{Q} =
\begin{bmatrix}
\mat{Q}_1           \\
& \ddots           \\
&& \mat{Q}_{N-1} \\
&&& \mat{Q}_{N}  \\
\end{bmatrix}\!\!,~
\end{align*}
\end{minipage}
}\\

\n with a constant term $c(\vec{x}_0) = \frac{1}{2}\vec{x}_0^T\mat{Q}_0\vec{x}_0 + \mat{q}_0^T\vec{x}_0$ which does not influence the minimizer. The system dynamics \eqref{eq:QP_equality_constraints} and box-constraints \eqref{eq:QP_inequality_constraints} are given by
\begin{align*}
\cmat{A} &=
\begin{bmatrix}
        \mat{I} &~          &~                  &~      \\
    -\mat{A}_1  &\mat{I}    &~                  &~      \\
    ~           &\ddots     &\ddots             &~      \\
    ~           &~          &-\mat{A}_{N-1}   &\mat{I}\\
\end{bmatrix}\!\!,~
\cmat{w}(\vec{x}_0) =
\begin{bmatrix}
    \mat{A}_0\vec{x}_0 + \vec{w}_0 \\ \vec{w}_1 \\ \vdots \\ \vec{w}_{N-1}
\end{bmatrix}\!\!,\\
\cmat{B} &=
\begin{bmatrix}
    \mat{B}_0   &~          &~      &~              \\
    ~           &\mat{B}_1  &~      &~              \\
    ~           &~          &\ddots &~              \\
    ~           &~          &~      &\mat{B}_{N-1}\\
\end{bmatrix}\!\!,~
\underline{\cmat{u}} =
\begin{bmatrix}
    \underline{\vec{u}}_0 \\ \underline{\vec{u}}_1 \\ \vdots \\ \underline{\vec{u}}_{N-1}
\end{bmatrix}\!\!,~
\overline{\cmat{u}} =
\begin{bmatrix}
    \overline{\vec{u}}_0 \\ \overline{\vec{u}}_1 \\ \vdots \\ \overline{\vec{u}}_{N-1}
\end{bmatrix}\!\!.~
\end{align*}
Notice that $\cmat{A}\in\mathbb{R}^{N\cdot n_\vec{x}\times N\cdot n_\vec{x}}$ is invertible by construction.

In the following, move-blocking and partial-condensing procedures are recalled. Then they are both incorporated in transformation in \secref{sec:Customization}, leading to the generalized QP formulation.

\section{Move Blocking}
\label{sec:Move-Blocking}
\ac{MB} is commonly used to deal with the computational burden in optimal control. The strategy is to fix an input or a change between two consecutive inputs to be constant for several time-steps \cite{Cagienard2007}. Thus, the number of degrees of freedom in the optimization problem is reduced significantly for dense QP formulation.

The choice of the blocking strategy to provide robust control performance is provided in \cite{Shekhar2012}, where a generalized blocked variable-horizon \ac{MPC} is formulated. The optimal blocking strategy is proposed in \cite{Shekhar2015}. Therein, the optimality is measured regarding controller complexity and region of attraction volume and requires a solution to mixed-integer programming. Once a move-blocking strategy is chosen, however, it hasn't been shown how to decrease the degree of freedom in the optimization problem for sparse \ac{QP} formulation.

The level of blocking is parametrized by move-bloking vector $\vec{m}_\mathrm{MB} = [m_1,m_2,\ldots,m_{N_\cvec{u}}]\in\mathbb{N}^{N_\cvec{u}}$, where $m$ are sizes of blocking windows and $N_\cvec{u}$ is number of input vectors after blocking to be optimized, and for which $\text{sum}(\vec{m}_\mathrm{MB}) = N$. For sake of brevity, an auxiliary vector of row indicies $\vec{j}_\mathrm{MB} = \textrm{cumsum}\footnote{Function \emph{cumsum} returns vector of cumulative sum of input argument. }(\vec{m}_\mathrm{MB}) = [j_1,j_2,\ldots,j_{N_\cvec{u}}]$ is defined. Then \emph{blocking} matrix $\cmat{T}$ and and input vectors after blocking $\blc{\cmat{u}}$ to be optimized are
\\\n\resizebox{\linewidth}{!}{
\begin{minipage}{\linewidth}
\begin{align*}
\cmat{T} =
\begin{bmatrix}
\mat{I} \\
\vdots \\
\mat{I}_{j_1} \\
& \mat{I} \\
& \vdots  \\
& \mat{I}_{j_2} \\
&& \ddots
\end{bmatrix}\in\mathbb{R}^{N\cdot n_\vec{u} \times N_\cvec{u}\cdot n_\vec{u}},\quad
\blc{\cmat{u}} =
\begin{bmatrix}
    \blc{\mat{u}}_{j_1} \\ \blc{\mat{u}}_{j_2} \\ \vdots \\ \blc{\mat{u}}_{j_{N_\vec{u}}}
\end{bmatrix}\in\mathbb{R}^{N_\cvec{u}\cdot n_\vec{u}},
\end{align*}
\end{minipage}
}\\

\n respectively. The blocking is provided by input transformation
\begin{equation}\label{eq:move-blocking}
\cmat{u} = \cmat{T}\blc{\cmat{u}}.
\end{equation}

\n Note that $\cmat{T}$ is matrix of ones and zeros only such that $\cmat{T}^{T}\cmat{T} = \mathrm{diag}({\vec{m}_\mathrm{MB}})\otimes\cmat{I}$. For the admissible $\cmat{T}$, an identity on each following row must be in the same column or the next right column. Move blocking \emph{approximates} the original problem with one with a lesser number of degrees of freedom. The approximation effect on the control performance has been discussed, e.g., in \cite{Schwickart2016}. Therein, the authors suggest a heuristic method to adapt the blocking strategy online such that control performance remains nearly unchanged. This paper focuses rather on computational and memory aspects.

In this paper, we modify \cite{Axehill2015} to allow the state condensing works for move-blocked \ac{MPC}.

\section{State condensing}
\label{sec:Partial-Condensing}
The state condensing was introduced in \cite{Axehill2015}. The idea is to eliminate not every, but only some of the state vectors from along the prediction horizon. This will result in an optimization problem where equality constraints representing system dynamics are eliminated out only partially. The idea is to take advantage of both sparse and dense \ac{QP} formulation, as in the partially-condensed problem, some sparsity structure remains, and simultaneously, the number of variables is reduced.

The level of condensing is parametrized by a state condensing vector $\vec{p}_\mathrm{PC} = [p_1,p_2,\ldots,p_{N_\cvec{x}}]\in\mathbb{N}^{N_\cvec{x}}$, where $p$ are sizes of condensing windows and $N_\cvec{x}$ is number of states vectors after condensing. For sake of brevity, a vector of row indicies $\vec{i}_\mathrm{PC} = \textrm{cumsum}(\vec{p}_\mathrm{PC}) = [i_1,i_2,\ldots,i_{N_\cvec{x}}]$ is defined. Possible option is $N_\cvec{x}=0$, i.e. $\vec{p}_\mathrm{PC}$ is empty vector which results in dense formulation. Then \emph{condensing} matrix and state vectors after condensing to be optimized are
\\\n\resizebox{\linewidth}{!}{
\begin{minipage}{\linewidth}
\begin{align*}
\cmat{E} &=
\begin{bmatrix}
\mat{I}_{i_1} \\
\mat{0} \\
\vdots \\
& \mat{I}_{i_2} \\
& \mat{0} \\
& \vdots  \\
&& \ddots
\end{bmatrix}\in\mathbb{R}^{N\cdot n_\vec{x} \times N_\cvec{x}\cdot n_\vec{x}},
\quad
\blc{\cmat{x}} =
\begin{bmatrix}
    \blc{\mat{x}}_{i_1} \\ \blc{\mat{x}}_{i_2} \\ \vdots \\ \blc{\mat{x}}_{i_{N_\vec{x}}}
\end{bmatrix}\in\mathbb{R}^{N_\cvec{x}\cdot n_\vec{x}},
\end{align*}
\end{minipage}
}\\

\n respectively. The remaining states $\blc{\cmat{x}}$ and states being condensed out $\rem{\cmat{x}}$ are separable as 
\begin{align}
\cmat{E}\blc{\cmat{x}} + \cmat{F}\rem{\cmat{x}} &= \cmat{x}, \cmat{E}^T\cmat{F}=\cmat{F}^T\cmat{E}=\cmat{0} \implies \cmat{E}\cmat{E}^T\cmat{x}=\cmat{E}\blc{\cmat{x}} \label{eq:sel}
\end{align}
where
\\\n\resizebox{\linewidth}{!}{
\begin{minipage}{\linewidth}
\begin{align*}
\cmat{F} &=
\begin{bmatrix}
\vdots \\
\mat{I}_{\rem{i}_1} \\
& \ddots \\
&& \mat{I}_{\rem{i}_1} \\
&& \mat{0} \\
&&&\mat{I}_{\rem{i}_2} \\
&&&& \ddots \\
&&&&& \mat{I}_{\rem{i}_2} \\
&&&&& \mat{0} \\
&&&&&& \ddots
\end{bmatrix}\in\mathbb{R}^{N\cdot n_\vec{x} \times (N-N_\cvec{x})\cdot n_\vec{x}},
\quad
\rem{\cmat{x}} =
\begin{bmatrix}
    \rem{\mat{x}}_{\rem{i}_1} \\ \rem{\mat{x}}_{\rem{i}_2} \\ \vdots \\ \rem{\mat{x}}_{\rem{i}_{N-N_\vec{x}}}
\end{bmatrix}\in\mathbb{R}^{(N-N_\cvec{x})\cdot n_\vec{x}},
\end{align*}
\end{minipage}
}\\

The prediction \eqref{eq:QP_equality_constraints} can be decomposed on the rows related to the remaining states using $\cmat{E}\cmat{E}^T$ and the rest using $\cmat{F}\cmat{F}^T$, respectively, as
\begin{align}
\cmat{E}\cmat{E}^T\cmat{A}\cmat{x} &= \cmat{E}\cmat{E}^T\cmat{B}\cmat{u} + \cmat{E}\cmat{E}^T\cmat{w}(\vec{x}_0), \label{eq:selPred} \\
\cmat{F}\cmat{F}^T\cmat{A}\cmat{x} &= \cmat{F}\cmat{F}^T\cmat{B}\cmat{u} + \cmat{F}\cmat{F}^T\cmat{w}(\vec{x}_0). \label{eq:notselPred}
\end{align}
Adding \eqref{eq:sel} to \eqref{eq:notselPred} a partial state prediction can be written down
\begin{align}\label{eq:partialPrediction}
(\cmat{E}\cmat{E}^T + \cmat{F}\cmat{F}^T\cmat{A})\cmat{x} &= \cmat{E}\blc{\cmat{x}} + \cmat{F}\cmat{F}^T\cmat{B}\cmat{u} + \cmat{F}\cmat{F}^T\cmat{w}(\vec{x}_0),\nonumber\\
\text{or}\qquad\qquad\cmat{x} &= \cmat{M}^{-1}\cmat{E}\blc{\cmat{x}} + \cmat{N}\cmat{u} + \cmat{b},
\end{align}
where
\begin{align*}
\cmat{M} &= \cmat{E}\cmat{E}^T + \cmat{F}\cmat{F}^T\cmat{A}, \\
\cmat{N} &= \cmat{M}^{-1}\cmat{F}\cmat{F}^T\cmat{B}, \\
\cmat{b} &= \cmat{M}^{-1}\cmat{F}\cmat{F}^T\cmat{w}(\vec{x}_0).
\end{align*}

The state condensing exploits banded structure of $\cmat{A}$. For the sake of brevity, the structure of the $\cmat{M}$ matrix follows 
\\\n\resizebox{\linewidth}{!}{
\begin{minipage}{\linewidth}
\begin{align*}
\cmat{M} &=
\begin{bmatrix}
{\cmat{A}}_{0,p_1} \\
& {\cmat{A}}_{i_1,p_2} \\
&& \ddots \\
\end{bmatrix}\!\!,~
{\cmat{A}}_{i,p} =
\begin{bmatrix}
\cmat{I} \\
\cmat{A}_{i+1} & \cmat{I} \\
& \ddots & \ddots \\
&& \cmat{A}_{i+p} & \cmat{I}
\end{bmatrix}\!\!.
\end{align*}
\end{minipage}
}\\
Note that $\cmat{M}$ remains invertible, moreover, $\cmat{M}^{-1}$ is also unit lower triangular for any admissible $\cmat{E}$.
\\\n\resizebox{0.85\linewidth}{!}{
\begin{minipage}{\linewidth}
\begin{align}
\cmat{M}^{-1} &=
\begin{bmatrix}
{\cmat{A}}_{0,p_1}^{-1} \\
& {\cmat{A}}_{i_1,p_2}^{-1} \\
&& \ddots \\
\end{bmatrix}\!\!,~
{\cmat{A}}_{i,p}^{-1} =
\begin{bmatrix}
\cmat{I} \\
\cmat{A}_{i+1} & \cmat{I} \\
\vdots & \ddots & \ddots \\
\prod_{k=i+1}^{i+p}\cmat{A}_{k} & \cdots & \cmat{A}_{i+p} & \cmat{I}
\end{bmatrix}\!\!.
\end{align}
\end{minipage}
}\\

Substituting $\cmat{x}$ in \eqref{eq:selPred} by \eqref{eq:partialPrediction} leads to a new equality constraint
\begin{equation}
\cmat{E}^T\cmat{A}\cmat{M}^{-1}\cmat{E}\blc{\cmat{x}} = \cmat{E}^T(\cmat{B}-\cmat{A}\cmat{N})\cmat{u} + \cmat{E}^T(\cmat{w}(\vec{x}_0)-\cmat{A}\cmat{b}).
\end{equation}

\n Note that for fully sparse case $\cmat{E}=\cmat{I}, \cmat{F}\text{ is empty}$ implies $\cmat{M}, \cmat{N}, \cmat{b}$ are empty matrices of particular size, i.e. the transformation is not needed at all. On the other hand, for dense case $\cmat{E}\text{ is empty}, \cmat{F}=\cmat{I}$ implies $\cmat{M}=\cmat{A}, \cmat{N}=\cmat{A}^{-1}\cmat{B}, \cmat{b}=\cmat{A}^{-1}\cmat{w}(\vec{x}_0)$ and \eqref{eq:partialPrediction} yields an ordinary prediction. Further, only \eqref{eq:selPred} and \eqref{eq:partialPrediction} are going to be used.

State condensing transform the original problem into an \emph{equivalent} one of a smaller dimension. The state-condensing procedure benefits from the fact that $\cmat{M}$ is block-diagonal. Therefore, it can be computed block by block, and off-diagonal terms remain zero.

\section{Generalized QP Formulation}
\label{sec:Customization}
By interconnectiong both previous methods, namely move blocking \eqref{eq:move-blocking} and state condensing \eqref{eq:sel},\eqref{eq:partialPrediction}, a systematic transformation can be written down now
\begin{subequations}\label{eq:transform}
\begin{align}
\cmat{x} &= \cmat{\Upsilon}\blc{\cmat{x}} + \cmat{\Gamma}\blc{\cmat{u}} + \cmat{b}, \label{eq:transform1}\\
\cmat{u} &= \cmat{T}\blc{\cmat{u}}. \label{eq:transform2}
\end{align}
\end{subequations}
where $\cmat{\Upsilon} = \cmat{M}^{-1}\cmat{E}$ and $\cmat{\Gamma} = \cmat{N}\cmat{T}$.

The problem \eqref{eq:QP_problem} can be then transformed using \eqref{eq:transform} into a generalized \ac{QP} \eqref{eq:customizable_QP_problem} of the similar structure. The transformation \eqref{eq:transform} is applied on \eqref{eq:QP_problem} such that \eqref{eq:transform1} and \eqref{eq:transform2} are substitute in \eqref{eq:QP_problem}. Then the resulting generalized \ac{QP} problem is

\n\resizebox{\linewidth}{!}{
\begin{minipage}{\linewidth}
\begin{eqnarray}\label{eq:customizable_QP_problem}
& \underset{\blc{\cmat{u}},\blc{\cmat{x}}} {\min}& \dfrac{1}{2}
\begin{bmatrix}
\blc{\cmat{u}} \\
\blc{\cmat{x}}
\end{bmatrix}^T
\begin{bmatrix}
\blc{\cmat{R}} & \blc{\cmat{S}}^T \\
\blc{\cmat{S}} & \blc{\cmat{Q}}
\end{bmatrix}
\begin{bmatrix}
\blc{\cmat{u}} \\
\blc{\cmat{x}}
\end{bmatrix}
+
\begin{bmatrix}
\blc{\cmat{u}} \\
\blc{\cmat{x}}
\end{bmatrix}^T
\begin{bmatrix}
\blc{\cmat{r}}(\vec{x}_0) \\
\blc{\cmat{q}}(\vec{x}_0)
\end{bmatrix}
+ \blc{c}(\vec{x}_0)\nonumber\\
&  \textrm{s.t.}  & \blc{\cmat{A}}\blc{\cmat{x}} = \blc{\cmat{B}}\blc{\cmat{u}} + \blc{\cmat{w}}(\vec{x}_0), \nonumber\\
& & \underline{\blc{\cmat{u}}} \leq \blc{\cmat{u}} \leq \overline{\blc{\cmat{u}}},
\end{eqnarray}
\end{minipage}
}\\

\n where
\\\n\resizebox{0.88\linewidth}{!}{
\begin{minipage}{\linewidth}
\begin{align}\label{eq:customizable_QP_matrices}
\begin{split}
\blc{\cmat{R}}&=\cmat{T}^T\cmat{R}\cmat{T} + \cmat{\Gamma}^T(\cmat{Q}\cmat{\Gamma} + \cmat{S}\cmat{T}) + (\cmat{S}\cmat{T})^T\cmat{\Gamma},\\
\blc{\cmat{S}}&=\cmat{\Upsilon}^T(\cmat{Q}\cmat{\Gamma}+\cmat{S}\cmat{T}),\\
\blc{\cmat{Q}}&=\cmat{\Upsilon}^T\cmat{Q}\cmat{\Upsilon},\\
\blc{\cmat{A}}&=\cmat{E}^T\cmat{A}\cmat{\Upsilon},\\
\blc{\cmat{B}}&=\cmat{E}^T(\cmat{B}\cmat{T}-\cmat{A}\cmat{\Gamma}),\\
\overline{\blc{\cmat{u}}} &= \cmat{T}^+\overline{\cmat{u}}, \\
\underline{\blc{\cmat{u}}} &= \cmat{T}^+\underline{\cmat{u}},
\end{split}
\hspace{-8.5em}
\begin{split}
\\
\\
\blc{\cmat{r}}(\vec{x}_0) &= \cmat{T}^T\cmat{r}(\vec{x}_0) + (\cmat{S}\cmat{T})^T\cmat{b} + \cmat{\Gamma}^T(\cmat{q} + \cmat{Q}\cmat{b}), \\
\blc{\cmat{q}}(\vec{x}_0)&= \cmat{\Upsilon}^T(\cmat{q} + \cmat{Q}\cmat{b}), \\
\blc{\cmat{w}}(\mat{x}_0) &= \cmat{E}^T(\cmat{w}(\mat{x}_0)-\cmat{A}\cmat{b}),\\
\blc{c}(\mat{x}_0) &= c(\mat{x}_0) + 0.5\cmat{b}^T(\cmat{q} + \cmat{Q}\cmat{b}), \\
\blc{\cmat{u}}_0 &= \cmat{T}^+\cmat{u}_0, {}
\end{split}
\end{align}
\end{minipage}
}\\

\n and where $\cmat{T}^+$ denotes left-inverse of $\cmat{T}$. When the data for \eqref{eq:QP_problem} are available a generalized QP formulation \ac{QP} \eqref{eq:customizable_QP_problem} is given by \eqref{eq:customizable_QP_matrices} parametrized by $\cmat{T}$ and $\cmat{E}$. Matrices in \eqref{eq:customizable_QP_matrices} can be build efficiently with respect the sparsity structure and \ac{FLOPs} count can be obtained easily. Remember, any admissible choice of blocking matrix (except $\cmat{T} = \cmat{I}$) causes \eqref{eq:customizable_QP_problem} \emph{approximates} \eqref{eq:QP_problem} by problem with less degree of freedom. On the other hand, any admissible $\cmat{E}$ does not affect the minimization result.






Problem \eqref{eq:customizable_QP_problem} can be solved efficiently by an active-set or an interior-point method. Note that both $\mathrm{LDL}^\mathrm{T}$ and Cholesky decomposition typically used to find a Newton step within any of these methods preserves the sparsity pattern of this problem.

\section{Numerical Examples}
\label{sec:Numerical_Study}

For the sake of brevity, transformation matrices for a short prediction horizon problem setup are shown first. It is followed by a common benchmark -- the oscillating masses controlled by move-blocked MPC with a given blocking strategy. Finally, for a given system and prediction horizon, the whole transformation space is sampled to demonstrate the behavior of the proposed problem setups in more detail.

\subsection{Illustrative Example}
Assume a random \ac{LTV} system with ten states ($n_\vec{x}=10$) and five inputs ($n_\vec{u}=5$) for which sequence of move blocks was given to be $\vec{m}_\mathrm{MB} = [1,2,3]$, consequently, $\vec{j}_\mathrm{MB} = [1,3,6]$. Prediction horizon is $N=\text{sum}(\vec{m}_\mathrm{MB})=6$ and control horizon is $N_\cvec{u}=\text{length}(\vec{m}_\mathrm{MB})=3$. Let's choose $\vec{p}_\mathrm{PC} = [1,2,3]$ to be similar to the $\vec{m}_\mathrm{MB}$ in the first $N_\cvec{u}-1$ entries, consequently, $\vec{j}_\mathrm{PC} = [1,3,6]$. In this case, the transformation matrices in \eqref{eq:move-blocking} and \eqref{eq:sel} are given by
\begin{align*}
\cmat{T} &=
\begin{bmatrix}
\mat{I} \\
& \mat{I} \\
& \mat{I}\\
& & \mat{I} \\
& & \mat{I} \\
& & \mat{I}
\end{bmatrix}\!\!,~
\blc{\cmat{u}} =
\begin{bmatrix}
\blc{\vec{u}}_0 \\
\blc{\vec{u}}_1 \\
\blc{\vec{u}}_3
\end{bmatrix}\!\!,~
\cmat{E} =
\begin{bmatrix}
\mat{I} \\
& \mat{0} \\
& \mat{I}\\
&& \mat{0} \\
&& \mat{0} \\
&& \mat{I}
\end{bmatrix}\!\!,~
\blc{\cmat{x}} =
\begin{bmatrix}
\blc{\vec{x}}_1 \\
\blc{\vec{x}}_3 \\
\blc{\vec{x}}_6
\end{bmatrix}\!\!,
\\
\cmat{F} &=
\begin{bmatrix}
\mat{0} \\
\mat{I} \\
\mat{0}\\
& \mat{I} \\
&& \mat{I} \\
&& \mat{0}
\end{bmatrix}\!\!,~
\rem{\cmat{x}} =
\begin{bmatrix}
\rem{\vec{x}}_2 \\
\rem{\vec{x}}_4 \\
\rem{\vec{x}}_5
\end{bmatrix}\!\!,
\end{align*}
Consequently, condensing matrices in \eqref{eq:partialPrediction} and \eqref{eq:transform} are
\\\n\resizebox{\linewidth}{!}{
\begin{minipage}{\linewidth}
\begin{align*}
\cmat{M} &=
\begin{bmatrix}
\mat{I} \\
-\mat{A}_1  &\mat{I} \\
            &~          &\mat{I} \\
            &~          &-\mat{A}_3   &\mat{I} \\
            &~          &~            &-\mat{A}_4        &\mat{I} \\
            &~          &~            &~        &~       &\mat{I}
\end{bmatrix}\!\!,~
\cmat{N} =
\begin{bmatrix}
\mat{0} \\
            &\mat{B}_1 \\
            &           &\mat{0} \\
            &~          &          &\mat{B}_3 \\
            &~          &~         &\mat{A}_4\mat{B}_3  &\mat{B}_4 \\
            &~          &~         &~           &~         &\mat{0}
\end{bmatrix}\!\!,
\\
\cmat{M}^{-1} &=
\begin{bmatrix}
\mat{I} \\
\mat{A}_1 &\mat{I} \\
&& \mat{I} \\
&& \mat{A}_3 &\mat{I} \\
&& \mat{A}_4\mat{A}_3 & \mat{A}_4 & \mat{I} \\
&&&&& \mat{I}
\end{bmatrix}\!\!,~
\cmat{b} =
\begin{bmatrix}
\vec{0} \\
\vec{w}_2 \\
\vec{0} \\
\vec{w}_4 \\
\mat{A}_4\vec{w}_3 + \vec{w}_4 \\
\vec{0} \\
\end{bmatrix}\!\!.
\end{align*}
\end{minipage}
}\\

\n The sparsity structure of the resulting QP problem can be demonstrated at the \ac{KKT} matrix of QP problem \eqref{eq:customizable_QP_problem} with no inequality constraints being activated. The associated \ac{KKT} matrix is
\begin{equation}\label{eq:Hessian}
\blc{\cmat{H}} =
\begin{bmatrix}
\blc{\cmat{R}} & \blc{\cmat{S}}^T & \blc{\cmat{B}}^T \\
\blc{\cmat{S}} & \blc{\cmat{Q}} & \blc{\cmat{A}}^T \\
\blc{\cmat{B}} & \blc{\cmat{A}} & \cmat{0} 
\end{bmatrix}\!\!.
\end{equation}

\begin{figure}[ht]
\centering
\subfloat[Original]{%
\begin{tikzpicture}[scale=0.9]
\draw[->] (0,2.2) -- (0,4) node[anchor=north east] {$\vec{x}$};
\draw[->] (0,2.2) -- (6.5,2.2) node[anchor=north east] {};
\draw[->] (0,0) -- (6.5,0) node[anchor=north] {$t$};
\draw[->] (0,0) -- (0,2) node[anchor=north east] {$\vec{u}$};

\draw[-, dash pattern=on 3 pt off 3 pt] (0,4.1) -- (0,0) node[anchor=north] {$0$};
\draw[-, dash pattern=on 3 pt off 3 pt] (1,4.1) -- (1,0) node[anchor=north] {$1$};
\draw[-, dash pattern=on 3 pt off 3 pt] (2,4.1) -- (2,0) node[anchor=north] {$2$};
\draw[-, dash pattern=on 3 pt off 3 pt] (3,4.1) -- (3,0) node[anchor=north] {$3$};
\draw[-, dash pattern=on 3 pt off 3 pt] (4,4.1) -- (4,0) node[anchor=north] {$4$};
\draw[-, dash pattern=on 3 pt off 3 pt] (5,4.1) -- (5,0) node[anchor=north] {$5$};
\draw[-, dash pattern=on 3 pt off 3 pt] (6,4.1) -- (6,0) node[anchor=north] {$6$};

\draw[thick,red,-, dash pattern=on 10 pt off 5 pt] (0,0.4) -- (6,0.4);
\draw[thick,red,-, dash pattern=on 10 pt off 5 pt] (0,1.8) -- (6,1.8);
\draw[thick,-*] (1,0.4) -- (-.1,0.4) node[anchor=south west] {$\vec{u}_{0}$};
\draw[thick,-] (1,0.4) -- (1,1.7);
\draw[thick,-*] (2,1.7) -- (0.9,1.7) node[anchor=north west] {$\vec{u}_{1}$};
\draw[thick,-] (2,1.7) -- (2,1.8);
\draw[thick,-*] (3,1.8) -- (1.9,1.8) node[anchor=north west] {$\vec{u}_{2}$};
\draw[thick,-] (3,1.8) -- (3,0.4);
\draw[thick,-*] (4,0.4) -- (2.9,0.4) node[anchor=south west] {$\vec{u}_{3}$};
\draw[thick,-] (4,0.4) -- (4,0.9);
\draw[thick,-*] (5,0.9) -- (3.9,0.9) node[anchor=north west] {$\vec{u}_{4}$};
\draw[thick,-] (5,0.9) -- (5,1.0);
\draw[thick,-*] (6,1.0) -- (4.9,1.0) node[anchor=north west] {$\vec{u}_{5}$};

\coordinate (x0) at (0,2.8);
\coordinate (x1) at (1,2.6);
\coordinate (x1s) at (1,2.8);
\coordinate (x2) at (2,3.0);
\coordinate (x2s) at (2,2.8);
\coordinate (x3) at (3,3.6);
\coordinate (x3s) at (3,3.3);
\coordinate (x4) at (4,3.7);
\coordinate (x4s) at (4,3.4);
\coordinate (x5) at (5,3.6);
\coordinate (x5s) at (5,3.4);
\coordinate (x6) at (6,3.3);
\coordinate (x6s) at (6,3.0);

\draw [thick,black] (x0) to[out=50,in=180] (x1);
\draw [thick,black] (x1s) to[out=0,in=200] (x2);
\draw [thick,black] (x2s) to[out=50,in=200] (x3);
\draw [thick,black] (x3s) to[out=0,in=180] (x4);
\draw [thick,black] (x4s) to[out=0,in=180] (x5);
\draw [thick,black] (x5s) to[out=0,in=130] (x6);

\node [label=west:      {$\vec{x}_{0}$} ,fill=white,circle,draw,inner sep=0pt,minimum width=0.2cm] at (x0) {}; 
\node [fill=white,circle,draw,inner sep=0pt,minimum width=0.2cm] at (x1) {};
\node [fill=white,circle,draw,inner sep=0pt,minimum width=0.2cm] at (x2) {};
\node [fill=white,circle,draw,inner sep=0pt,minimum width=0.2cm] at (x3) {};
\node [fill=white,circle,draw,inner sep=0pt,minimum width=0.2cm] at (x4) {};
\node [fill=white,circle,draw,inner sep=0pt,minimum width=0.2cm] at (x5) {};
\node [fill=white,circle,draw,inner sep=0pt,minimum width=0.2cm] at (x6) {};

\node [label=north east:{$\vec{x}_{1}$} ,fill=black,circle,draw,inner sep=0pt,minimum width=0.2cm] at (x1s) {};
\node [label=east:      {$\vec{x}_{2}$} ,fill=black,circle,draw,inner sep=0pt,minimum width=0.2cm] at (x2s) {};
\node [label=south east:{$\vec{x}_{3}$} ,fill=black,circle,draw,inner sep=0pt,minimum width=0.2cm] at (x3s) {};
\node [label=south east:{$\vec{x}_{4}$} ,fill=black,circle,draw,inner sep=0pt,minimum width=0.2cm] at (x4s) {};
\node [label=south east:{$\vec{x}_{5}$} ,fill=black,circle,draw,inner sep=0pt,minimum width=0.2cm] at (x5s) {};
\node [label=east:      {$\vec{x}_{6}$} ,fill=black,circle,draw,inner sep=0pt,minimum width=0.2cm] at (x6s) {};
\end{tikzpicture}
}\\
\subfloat[Transformed]{%
\begin{tikzpicture}[scale=0.9]
\draw[->] (0,2.2) -- (0,4) node[anchor=north east] {$\blc{\vec{x}}$};
\draw[->] (0,2.2) -- (6.5,2.2) node[anchor=north east] {};
\draw[->] (0,0) -- (6.5,0) node[anchor=north] {$t$};
\draw[->] (0,0) -- (0,2) node[anchor=north east] {$\blc{\vec{u}}$};

\draw[-, dash pattern=on 3 pt off 3 pt] (0,4.1) -- (0,0) node[anchor=north] {$0$};
\draw[-, dash pattern=on 3 pt off 3 pt] (1,4.1) -- (1,0) node[anchor=north] {$1$};
\draw[-, dash pattern=on 3 pt off 3 pt] (2,4.1) -- (2,0) node[anchor=north] {$2$};
\draw[-, dash pattern=on 3 pt off 3 pt] (3,4.1) -- (3,0) node[anchor=north] {$3$};
\draw[-, dash pattern=on 3 pt off 3 pt] (4,4.1) -- (4,0) node[anchor=north] {$4$};
\draw[-, dash pattern=on 3 pt off 3 pt] (5,4.1) -- (5,0) node[anchor=north] {$5$};
\draw[-, dash pattern=on 3 pt off 3 pt] (6,4.1) -- (6,0) node[anchor=north] {$6$};

\draw[thick,red,-, dash pattern=on 10 pt off 5 pt] (0,0.4) -- (6,0.4);
\draw[thick,red,-, dash pattern=on 10 pt off 5 pt] (0,1.8) -- (6,1.8);
\draw[thick,-*] (1,0.4) -- (-.1,0.4) node[anchor=south west] {$\blc{\vec{u}}_{0}$};
\draw[thick,-] (1,0.4) -- (1,1.7);
\draw[thick,-*] (2,1.7) -- (0.9,1.7) node[anchor=north west] {$\blc{\vec{u}}_{1}$};
\draw[thick,-] (3,1.7) -- (2,1.7) node[fill=white,circle,draw,inner sep=0pt,minimum width=0.2cm] {};
\draw[thick,-] (3,1.7) -- (3,1.0);
\draw[thick,-*] (4,1.0) -- (2.9,1.0) node[anchor=north west] {$\blc{\vec{u}}_{3}$};
\draw[thick,-] (5,1.0) -- (4,1.0) node[fill=white,circle,draw,inner sep=0pt,minimum width=0.2cm] {};
\draw[thick,-] (6,1.0) -- (5,1.0) node[fill=white,circle,draw,inner sep=0pt,minimum width=0.2cm] {};

\coordinate (u2) at (2,1.7);
\coordinate (u4) at (4,1);
\coordinate (u5) at (5,1);


\coordinate (x0) at (0,2.8);
\coordinate (x1) at (1,2.6);
\coordinate (x1s) at (1,2.8);
\coordinate (x2) at (2,3.2);
\coordinate (x3) at (3,3.7);
\coordinate (x3s) at (3,3.3);
\coordinate (x4) at (4,3.6);
\coordinate (x5) at (5,3.7);
\coordinate (x6) at (6,3.5);
\coordinate (x6s) at (6,3.0);

\draw [thick,black] (x0) to[out=50,in=180] (x1);
\draw [thick,black] (x1s) to[out=0,in=200] (x2);
\draw [thick,black] (x2) to[out=30,in=180] (x3);
\draw [thick,black] (x3s) to[out=0,in=180] (x4);
\draw [thick,black] (x4) to[out=0,in=180] (x5);
\draw [thick,black] (x5) to[out=0,in=130] (x6);

\node [label=west:      {$\vec{x}_{0}$} ,fill=white,circle,draw,inner sep=0pt,minimum width=0.2cm] at (x0) {}; 
\node [fill=white,circle,draw,inner sep=0pt,minimum width=0.2cm] at (x1) {};
\node [fill=white,circle,draw,inner sep=0pt,minimum width=0.2cm] at (x2) {};
\node [fill=white,circle,draw,inner sep=0pt,minimum width=0.2cm] at (x3) {};
\node [fill=white,circle,draw,inner sep=0pt,minimum width=0.2cm] at (x4) {};
\node [fill=white,circle,draw,inner sep=0pt,minimum width=0.2cm] at (x5) {};
\node [fill=white,circle,draw,inner sep=0pt,minimum width=0.2cm] at (x6) {};

\node [label=north east:{$\blc{\vec{x}}_{1}$} ,fill=black,circle,draw,inner sep=0pt,minimum width=0.2cm] at (x1s) {};
\node [label=south east:{$\blc{\vec{x}}_{3}$} ,fill=black,circle,draw,inner sep=0pt,minimum width=0.2cm] at (x3s) {};
\node [label=east:      {$\blc{\vec{x}}_{6}$} ,fill=black,circle,draw,inner sep=0pt,minimum width=0.2cm] at (x6s) {};
\end{tikzpicture}
}
\caption{Sketch of the optimization problem. Comparison of the original and transformed problem. {\LARGE$\bullet$} denotes vectors included in and {\LARGE$\circ$} vectors excluded from the optimization.}
\label{fig:problemReduction}
\end{figure}
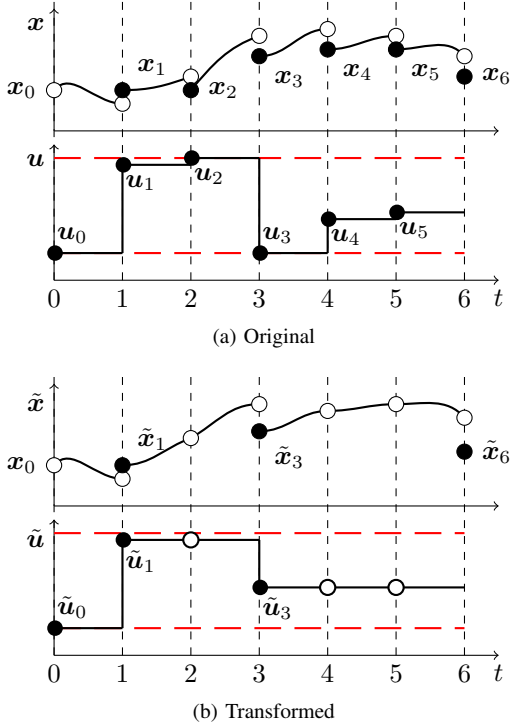

\begin{figure}[ht]
\hspace*{-0.5cm}
\subfloat[Original]{%
\includegraphics[width=0.6\linewidth]{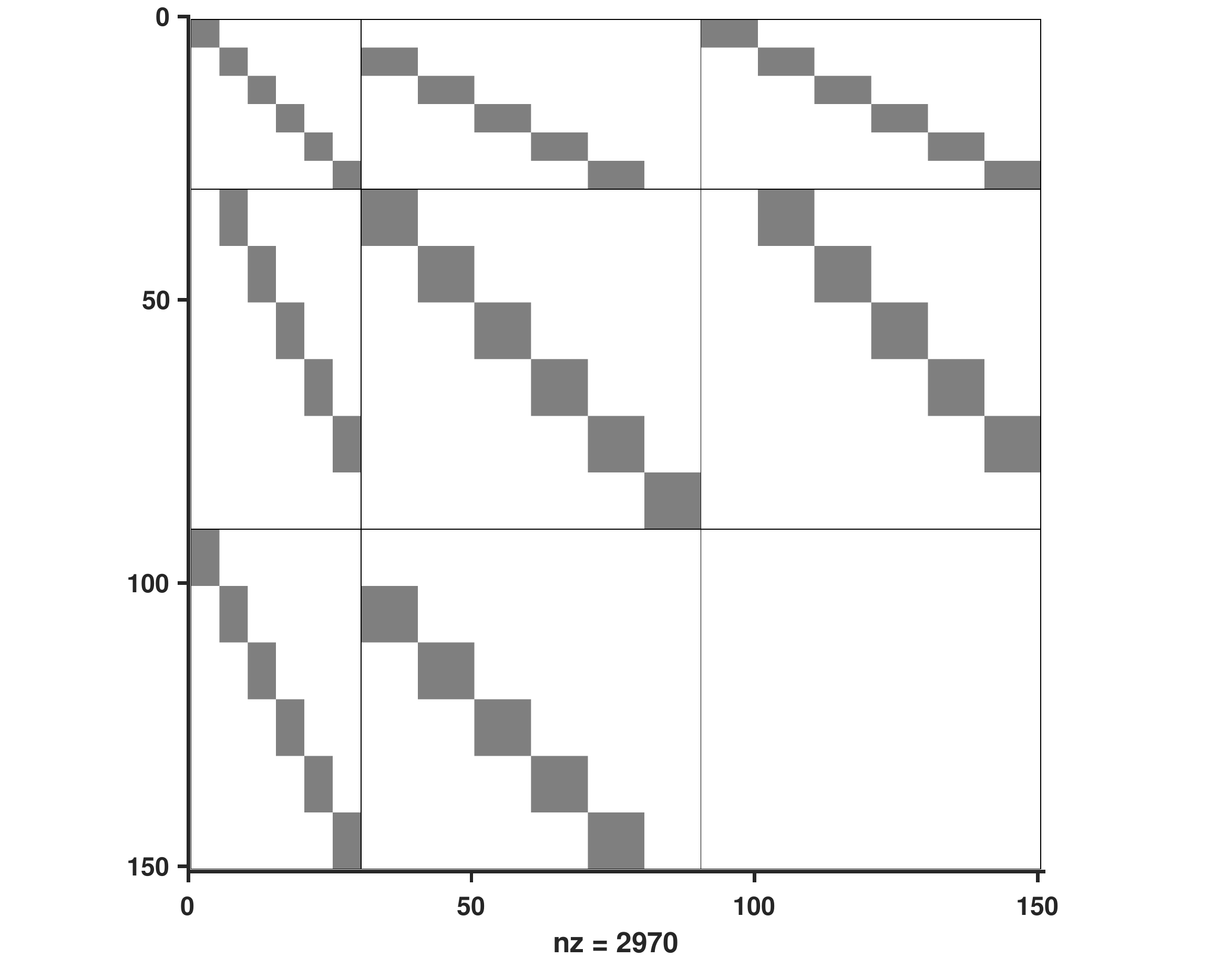}
}
\hspace*{-0.75cm}
\subfloat[Transformed]{%
\includegraphics[width=0.6\linewidth]{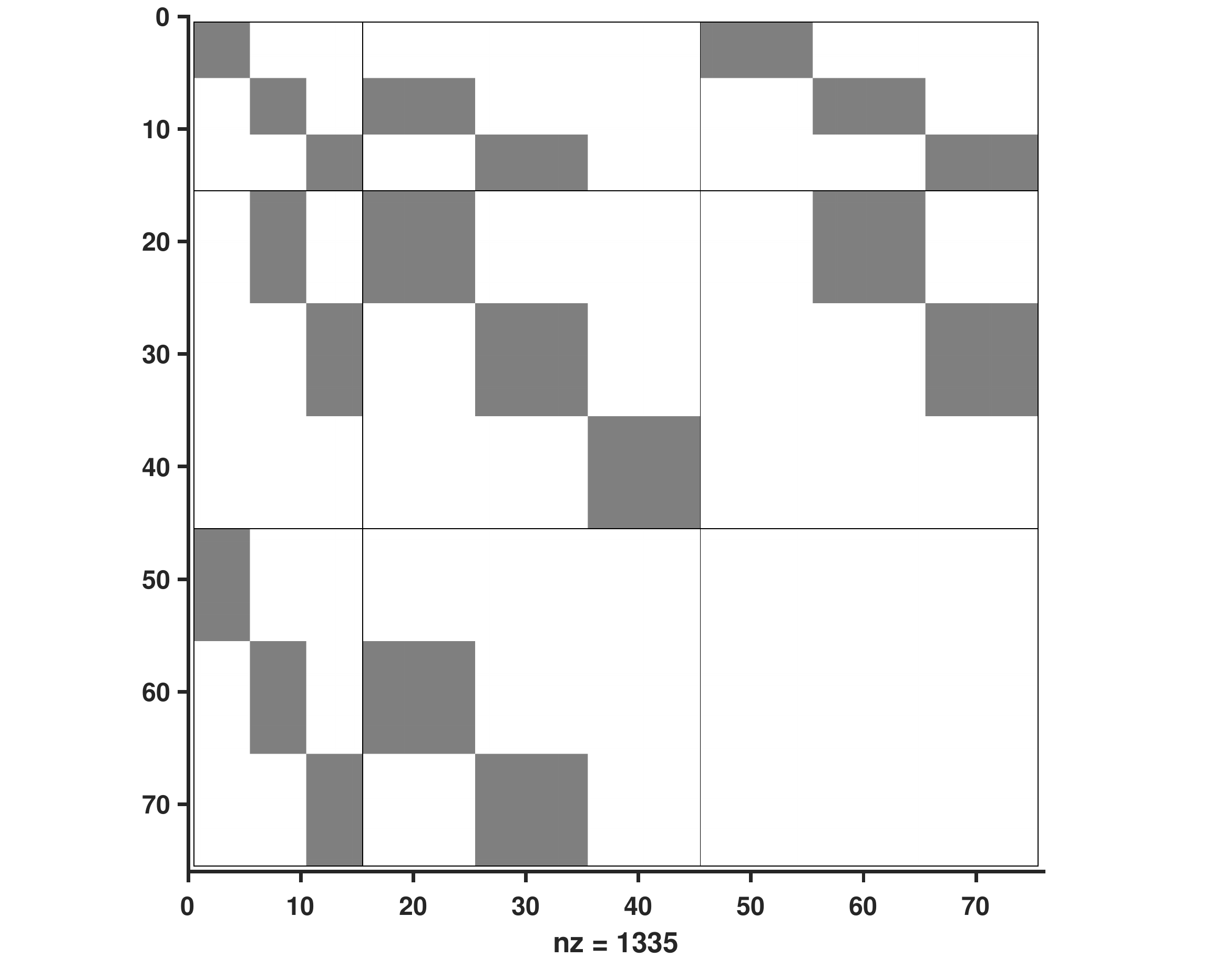}
}
\caption{Sparsity patterns of the \ac{KKT} matrix $\tilde{\cmat{H}}$ associated with a short prediction horizon example. All inequality constraints are assumed to be inactive.}
\label{fig:example-small}
\end{figure}

The example is sketched in \figref{fig:problemReduction} and sparsity pattern of the associated \ac{KKT} matrix \eqref{eq:Hessian} in \figref{fig:example-small}.

This simple example illustrates (see \figref{fig:example-small}) that the structure pattern of the problem is invariant for the proposed transformation. Also notice, the transformed problem has a smaller dimension, and less than half non-zero elements; therefore, it is expected the transformed problem requires less computational effort to a solution.

\subsection{Oscillating Masses} \label{sec:oscillating_masses}
This benchmark is inspired by \cite{Wang2008}. The proposed system consists of a sequence of six masses connected {to each other by spring dampers}. The first and the last masses are connected to the walls. The weight of {each mass} is $1$ {kg} and {the} spring constant is $1$ {N/m} without damping. The system state $\mat{x} \in \mathbb{R}^{12}$ represents the displacement and velocity of an individual mass. There are four control inputs, i.e., $\mat{u} \in \mathbb{R}^4$, which exert tensions between different masses. We assume control limits $-0.5 \leq \mat{u} \leq 0.5$, and the presence of random bounded external disturbance $\mat{v} \in \mathbb{R}^{6}$ with a uniform distribution on $\left[-0.5, 0.5\right]${,} which acts additionally on the displacement state of each mass. The control objective is to stabilize each mass in its origin, i.e., to solve \eqref{eq:MPC_problem} with $\mat{Q}=\mat{I}$, $\mat{R} = \mat{I}$, $\mat{S} = \mat{0}$, $\mat{q} = \mat{0}$, $\mat{r} = \mat{0}$, and with sampling time $T_s = 0.5$ s.



Further, we focus on the computational cost of problem \eqref{eq:customizable_QP_problem} build (preparation phase) and its solution (feedback phase). This is typically studied in nonlinear MPC, where \eqref{eq:customizable_QP_problem} has to be built and solved every sampling period \cite{Diehl2005}. The more QP problem is condensed and/or blocked, the more expansive the preparation is. On the other hand, the preparation may sufficiently decrease the solution time in particular. The preparation and feedback phase for some specific cases are examined numerically, and \ac{FLOPs} are measured.

The feedback phase denotes the cost of problem \eqref{eq:customizable_QP_problem} solution, in this paper, it is counted for \emph{NPPsparse} solver \cite{NPPsparse}. This active-set-like method converges typically in several iterations. The method benefits from the use of warm/hot-start while the number of iterations is insensitive to the problem conditioning.

\begin{figure}[ht]
\centering
\begin{tikzpicture}[scale=1]
\node[inner sep=0pt] (russell) at (0,0)
{
\subfloat[Computational requirements -- number of \ac{FLOPs} of NPPsparse \cite{NPPsparse}.]{
\includegraphics[width=0.9\linewidth]{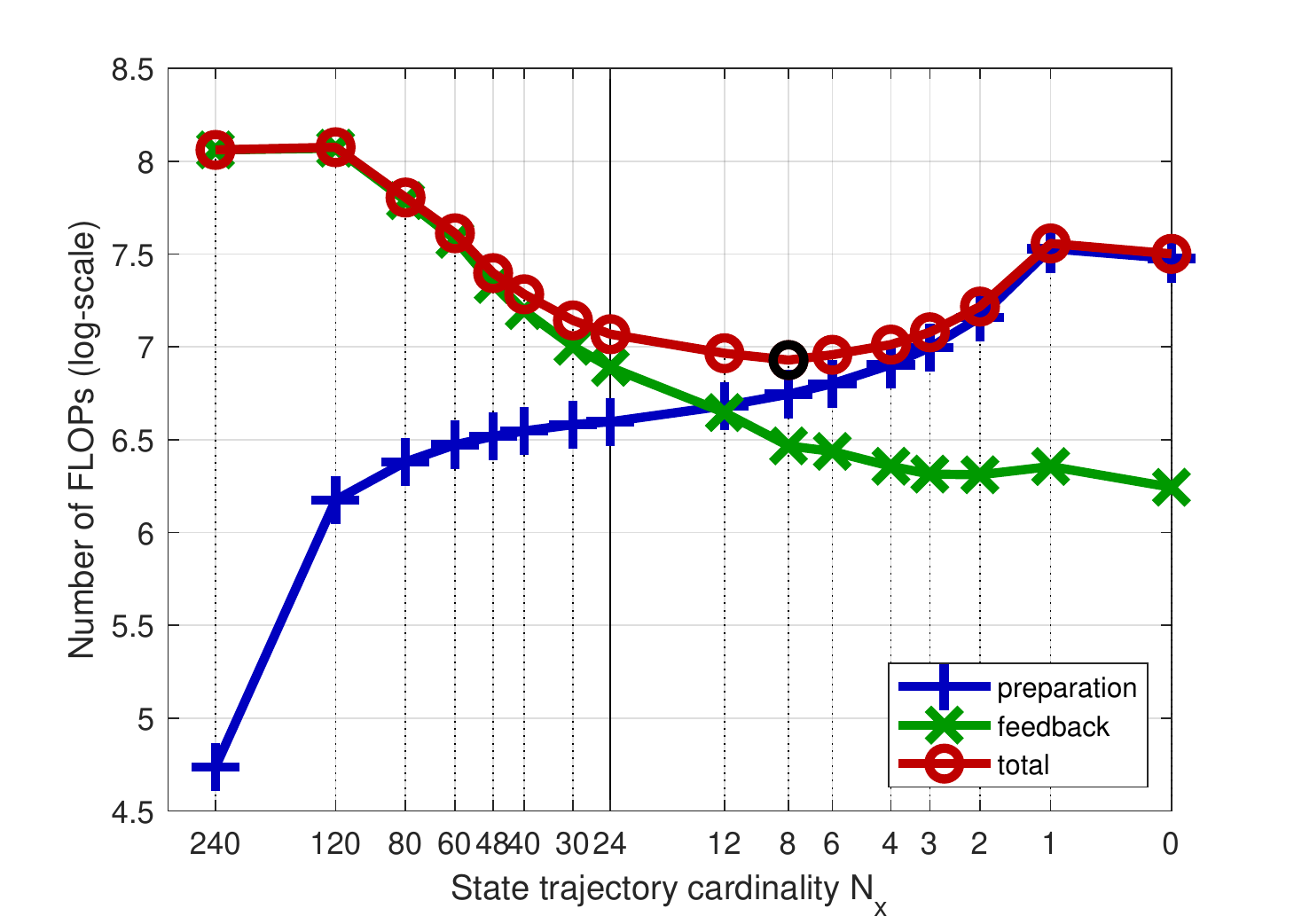}
}};
\end{tikzpicture}
\\
\subfloat[Memory requirements -- number of nonzero elements in \eqref{eq:Hessian}.]{%
\includegraphics[width=0.9\linewidth]{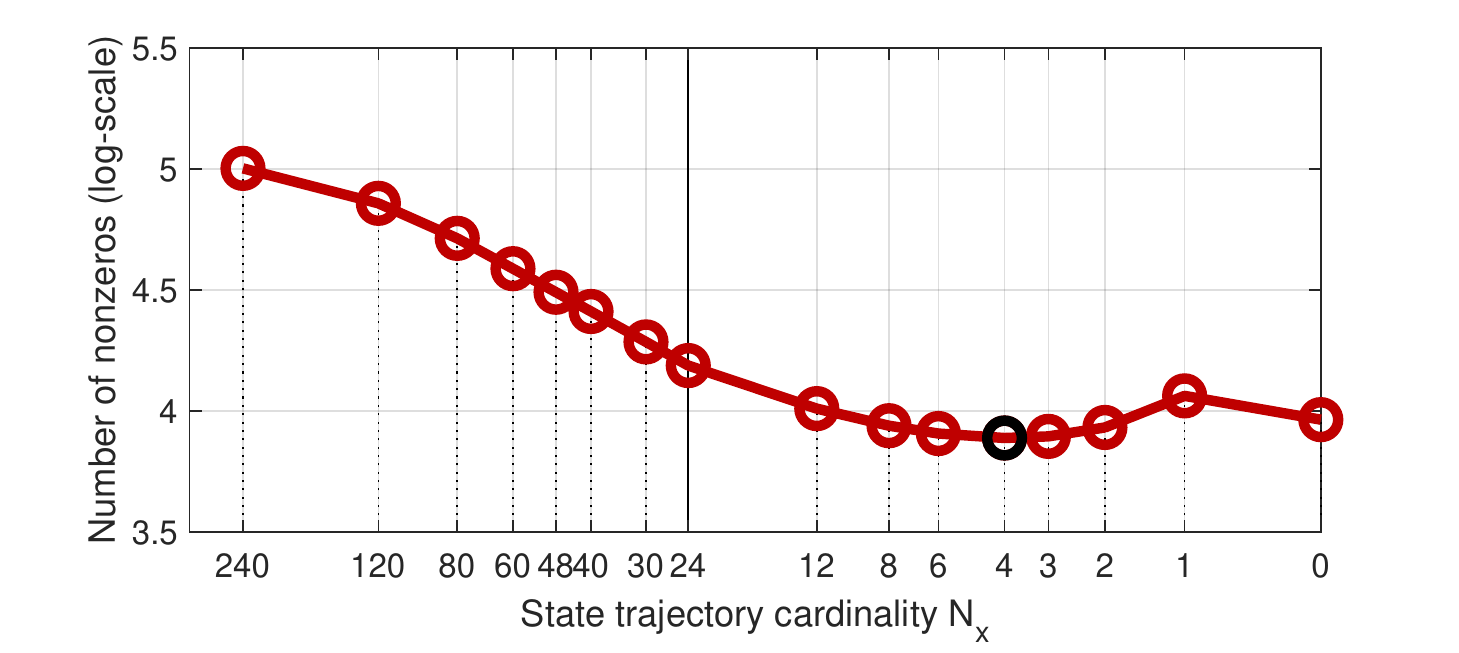}
}
\caption{Computational and memory requirements for equivalent \ac{QP} representing different level of condensing of move-blocked \ac{MPC} for oscillating masses. {\Large$\circ$} denote optimal requirements.}
\label{fig:results-oscillating_masses}
\end{figure}

In this exemplary case, the move-blocking strategy was chosen to be
$$\vec{m}_\mathrm{MB} = [10,\ldots,10]\in\mathbb{N}^{24}.$$
Once the blocking strategy is fixed, the control performance is immutable, i.e., all examined \ac{QP}s are equivalent. For this setup, various levels of condensing are tested. Condensing vector is chosen such that any state vector is closing move-blocking series, more precisely
$$
\vec{p}_\mathrm{PC} =
\left\{
\begin{aligned}
& \mathrm{empty~vector},~ i=0,\\
& [240/i,\ldots,240/i]\in\mathbb{N}^{i},\\ &\qquad\qquad i\in\{1,2,3,4,5,6,8,12,15,16,20,24\},
\end{aligned}
\right.
$$
and
$$\vec{p}_\mathrm{PC} = [10,\ldots,10]\in\mathbb{N}^{240/i}, i\in\{30,40,48,60,80,120,240\}$$
where in the later some of the block windows are additionally splitted in half to obtain the proper length $i$ of vector $\vec{p}_\mathrm{PC}$. Resulting computational and memory requirements for this setup are shown in \figref{fig:results-oscillating_masses}.

In \figref{fig:results-oscillating_masses}, on the left ($N_\cmat{x} = 240$) is the sparse \ac{QP} and on the right ($N_\cmat{x} = 0$) is the dense problem formulation. The first observation is that minimal computational or memory burden is in between these two. In other words, it is beneficial to condense the original sparse \ac{QP} partially. Specifically, computational requirements of the optimally condensed ($\approx 10^{7}$\ac{FLOPs}) compared to the dense \ac{QP} ($\approx 10^{7.5}$\ac{FLOPs}) is more than three times lower. One can save $20\%$ of memory space in case of optimally condensed ($\approx 10^{3.9}$\ac{NNZs}) compared to the dense \ac{QP} ($\approx 10^{4}$\ac{NNZs}). Another observation is that preparation cost when no condensing is required is significantly smaller compared to any other level of condensing.

In general, the computational cost of the preparation phase grows with the level of condensing. On the other hand, the computational cost of the feedback phase is not monotone and changes depending on multiple factors ($N_\cvec{x} ,N_\cvec{u},n_\vec{x},n_\vec{u}$) and importantly, on a particular algorithm implementation.


The presented numerical experience illustrates that using generalized QP formulation \eqref{eq:customizable_QP_problem}, for a given move-blocked \ac{MPC}, an optimization problem that requires minimal \ac{FLOPs} can be found.

The proposed approach has been implemented and tested in MATLAB environment, which allows code generation to an embedded platform. There is no need for two separate pieces of code (for dense and sparse \ac{QP} formulation), which labor-saving of code maintenance.

\section{Conclusion}
In this paper, we combine move-blocking and state-condensing procedures in generalized QP formulation of a move-blocked MPC problem. The combination of move-blocking and state-condensing methods allows reducing the number of input variables as well as a number of state variables. This approach allows for a \ac{MPC} with a given move-blocking strategy to find such a QP formulation for which a total computational burden or memory footprint of the MPC regulator is minimal. It has been illustrated how the proposed transformation affects memory footprint and computational burden by numerical examples. We analyzed on the example, the computational burden could be significantly decreased ($\approx 3\times$) or memory saved ($\approx 20\%$). The proposed approach requires specialized \ac{QP} solver used together with an optimized library for sparse linear algebra.

\bibliography{On_the_QP_Solution_for_MPC_with_Move_Blocking}
\bibliographystyle{plain}

\end{document}